\documentclass[]{aastex631} 
\usepackage{amsmath}

\usepackage{xcolor} 
 
\usepackage{float} 
\usepackage{graphicx}
\DeclareUnicodeCharacter {2212} {-} 
 
\def\VLSR  {$V_{\rm LSR}$} 
\def \kms {km s$^{-1}$} 
 
\shorttitle{Parallax and 3D kinematics of water masers in G034.43+0.24} 
\shortauthors{Mai et al.} 
 
\graphicspath{{./}{figures/}}

\begin{document}

\title{The parallax and 3D kinematics of water masers in the massive
star-forming region G034.43+0.24}

\author[0000-0001-7573-0145]{Xiaofeng Mai} 
\affiliation{Shanghai Astronomical Observatory, Chinese Academy 
of Sciences, Shanghai 200030, People’s Republic of China}             
\affiliation{School of Astronomy and Space Sciences, University of
Chinese Academy of Sciences, No. 19A Yuquan Road, Beijing 100049,
People’s Republic of China}                                           
 
\author[0000-0003-1353-9040]{Bo Zhang} 
\affiliation{Shanghai Astronomical Observatory, Chinese Academy 
of Sciences, Shanghai 200030, People’s Republic of China}             
 
\author[0000-0001-7223-754X]{M.J. Reid} 
\affiliation{Center for Astrophysics $\vert$ Harvard \& Smithsonian,
60 Garden Street, Cambridge, MA 02138, USA}                             
 
\author[0000-0002-8517-8881]{L. Moscadelli} 
\affiliation{INAF–Osservatorio Astrofisico di Arcetri, Largo E.Fermi
5,50125 Firenze, Italy}                                                 
 
\author[0000-0003-2953-6442]{Shuangjing Xu} 
\affiliation{Shanghai Astronomical Observatory, Chinese Academy 
of Sciences, Shanghai 200030, People’s Republic of China}             
\affiliation{Korea Astronomy and Space Science Institute, 776 Daedeok-daero,
Yuseong-gu, Daejeon 34055, Republic of Korea}                           
 
\author[0000-0002-8604-5394]{Yan Sun} 
\affiliation{Shanghai Astronomical Observatory, Chinese Academy 
of Sciences, Shanghai 200030, People’s Republic of China}             
 
\author{Jingdong Zhang} 
\affiliation{Shanghai Astronomical Observatory, Chinese Academy 
of Sciences, Shanghai 200030, People’s Republic of China}             
\affiliation{School of Astronomy and Space Sciences, University of
Chinese Academy of Sciences, No. 19A Yuquan Road, Beijing 100049,
People’s Republic of China}

\author[0000-0002-5519-0628]{Wen Chen} 
\affiliation{School of Astronomy and Space Sciences, University of
Chinese Academy of Sciences, No. 19A Yuquan Road, Beijing 100049,
People’s Republic of China}                                           
\affiliation{Yunnan Observatories, Chinese Academy of Sciences, Kunming
650216, Yunnan, People’s Republic of China}

\author{Shiming Wen} 
\affiliation{Shanghai Astronomical Observatory, Chinese Academy 
of Sciences, Shanghai 200030, People’s Republic of China}             
 
\author[0000-0003-4506-3171]{Qiuyi Luo} 
\affiliation{Shanghai Astronomical Observatory, Chinese Academy 
of Sciences, Shanghai 200030, People’s Republic of China}             
\affiliation{School of Astronomy and Space Sciences, University of
Chinese Academy of Sciences, No. 19A Yuquan Road, Beijing 100049,
People’s Republic of China}                                           
 
\author[0000-0001-6459-0669]{Karl M. Menten} 
\affiliation{Max-Plank-Institut für Radioastronomie, Auf dem Hügel
69, D-53121 Bonn, Germany}                                              
 
\author{Xingwu Zheng} 
\affiliation{School of Astronomy and Space Science, Nanjing University,
22 Hankou Road, Nanjing 210093, People’s Republic of China}           
 
\author[0000-0003-4468-761X]{Andreas Brunthaler} 
\affiliation{Max-Plank-Institut für Radioastronomie, Auf dem Hügel
69, D-53121 Bonn, Germany}                                              
 
\author{Ye Xu} 
\affiliation{Purple Mountain Observatory, Chinese Academy of Sciences,
Nanjing 210008, People’s Republic of China}                           
 
\author{Guangli Wang} 
\affiliation{Shanghai Astronomical Observatory, Chinese Academy 
of Sciences, Shanghai 200030, People’s Republic of China}

\begin{abstract} 
 
We report a trigonometric parallax measurement of 22 GHz water masers
in the massive star-forming region G034.43+0.24 as part of the
Bar and Spiral Structure Legacy (BeSSeL) Survey using the Very Long
Baseline Array. The parallax is 0.330~$\pm$~0.018
mas, corresponding to a distance of 3.03$^{+0.17}_{-0.16}$ kpc. 
This locates G034.43+0.24 near the inner edge of the Sagittarius
spiral arm and at one end of a linear distribution of massive young
stars which cross nearly the full width of the arm.  The measured
3-dimensional motion of G034.43+0.24 indicates a near-circular Galactic
orbit.                                                                  
The water masers display arc-like distributions, possibly bow shocks,
associated with winds from one or more massive young stars.             
 
\end{abstract} 
 
\keywords{Astrometry --- Galaxy: fundamental parameters --- Galaxy:
kinematics and dynamics --- maser --- star: formation}

\section{Introduction} \label{sec:intro} 
 
    G034.43+0.24 is a well-studied infrared dark cloud (IRDC) which
hosts several massive protostars. Observations of its 1.2\,mm dust
emission reveal nine dense clumps (named MM1-9) which trace an elongated
filamentary structure (See Figure 1 in \citet{2006ApJ...651L.125W}).    
Methanol~\citep{2016ApJ...833...18H, 2017ApJS..231...20Y, 2019ApJS..241...18Y},
water~\citep{2001A&A...368..845V, 2006ApJ...651L.125W} and hydroxyl
masers~\citep{2014MNRAS.441.3137Q, 2019A&A...628A..90B} have been
detected in several clumps toward this IRDC, and the water maser        
G034.43+0.24 in this region, which is the target of this paper. 
 
    Previous 3\,mm continuum observations resolve G034.43+0.24 MM1
into seven dense cores (labeled MM1 a-g).  The masses
of the dense cores range from about 40 to 200 $M_{\sun}$\citep{2022MNRAS.510.5009L},
contributing to a total mass of MM1 of several hundreds of solar
masses and a bolometric luminosity of 3.2 $\times$ 10$^4 L_{\sun}$
within a diameter of 0.5 pc~\citep{2005ApJ...630L.181R}.                
\citet{2011ApJ...741..120R} reveal a complex kinematics of
MM1, including scale-dependent velocity gradients and six outflow lobes
associated with four independent outflows, which are seen in molecular
lines of HCO$^+$, CS, SiO, SO and CH$_3$OH by the Atacama Large
Millimeter/submillimeter Array's Three-millimeter Observation of Massive
Star-formation regions~\citep[ATOMS,][]{2020MNRAS.496.2790L, 2022MNRAS.510.5009L}
survey.                                                                 
 
    A kinematic distance for G034.43+0.24 of 3.7 kpc was estimated
from $^{13}$CO (1-0) \citep{2006ApJ...653.1325S} and CS (2-1) \citep{2004A&A...426...97F}
observations.  However, a trigonometric parallax of $0.643 \pm 0.049$ mas, corresponding
to a distance of $1.56^{+0.12}_{-0.11}$ kpc, was measured by \citet{2011PASJ...63..513K}
using the Japanese VLBI Exploration for Radio Astrometry (VERA) array.  
This parallax distance is less than half the kinematic distance,
and this discrepancy makes estimation of physical parameters of this
IRDC highly uncertain.                                                  
 
    The primary aim of this paper is to obtain an independent parallax
distance for G034.43+0.24 in order to resolve the current discrepancy,  
and accurately locate the dark cloud in the Galaxy. 
The Very Long Baseline Array (VLBA) observations presented here are part of the Bar
and Spiral Structure Legacy (BeSSeL) Survey\footnote{\url{http://bessel.vlbi-astrometry.org/}},
a Key Science Project of the National Radio Astronomy
Observatory\footnote{The National Radio Astronomy Observatory is a facility of
the National Science Foundation operated under cooperative agreement by Associated
Universities, Inc.}.
Here, we report a new parallax and proper motion for the water
masers associated with G34.43+0.24 MM1.                                 
In Section~\ref{sec:obs}, we describe our observations and data
reduction.  The parallax and proper motions are reported in Section~\ref{sec:result}.
In Section~\ref{sec:discuss}, we compare our parallax to the previous result, and briefly discuss the kinematics of the
water masers and the arm association. We summarize our results in Section~\ref{sec:sum}.        
 
\section{Observations and data reduction} 
\label{sec:obs} 
 
The parallax observations were conducted under VLBA program BR198O. 
We observed the 6$_{16}$-5$_{23}$ rotational transition of the H$_2$O
molecule (rest frequency 22.23508 GHz). The  observational setup
and calibration procedures follow those described in \citet{2009ApJ...693..397R}.
In this section, we outline details specific to this observation.       
Seven epochs (2013 Nov. 06, 2014 Apr. 10, Jul. 22, Sep. 20, Nov.
01, Dec. 06, and 2015 Apr. 10) were observed, each spanning 7 hr.       
Four adjacent 16 MHz bands of right and left circular polarization
were recorded. Spectral channels were                                   
spaced by 8 kHz, corresponding to a velocity of 0.11 km s$^{-1}$. 
We observed four strong (``fringe-finder'') calibrators, J1638+5720,
J1740+5211, J1824+1044, and J0019+7327,                                  
selected from the International Celestial Reference Frame 2 (ICRF2)
catalogue \citep{2015AJ....150...58F} in order to monitor all systems.  
Instrumental delay and phase differences among individual intermediate-frequency
(IF) bands were                                                         
calibrated with J1638+5720. 
 
\begin{figure}[ht] 
  \centering 
  \includegraphics[scale=0.53]{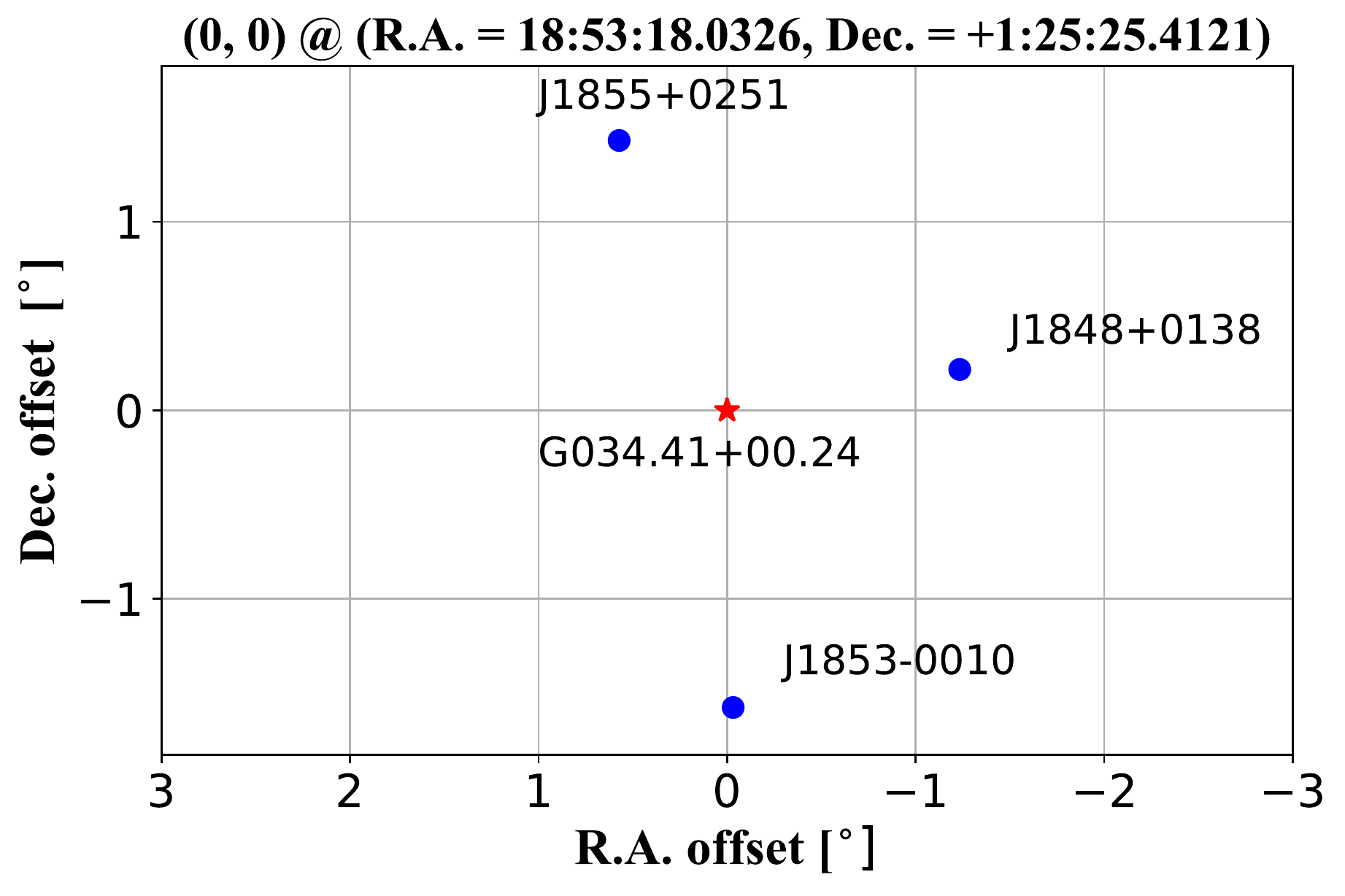} 
  \caption{ 
Sky distribution of the maser G034.43+0.24 and the
background quasars.                                                 
} 
  \label{fig:skydis} 
\end{figure}

The observations included four 0.5-hr ``geodetic blocks'' intermixed
with the                                                                
manual phase calibrator and phase-referencing observations.  The
geodetic blocks were used to calibrate and remove clock drifts and
uncompensated atmospheric delays~\citep{2009ApJ...693..397R}.           
Phase-reference blocks involved switching between G034.43+0.24 and
three background quasars                        
selected from the BeSSeL calibrator survey \citep{2011ApJS..194...25I}
in the following sequence:                                              
T,Q1,T,Q1,T,Q1, T,Q2,T,Q2,T,Q2, T,Q3,T,Q3,T,Q3 ...,
where T is the target G034.43+0.24, and Q1, Q2, Q3 are J1848+0138,
J1853$-$0010 and J1855+0251, respectively.
The locations of these sources are shown in Figure~\ref{fig:skydis}. 
Individual scans were 32 s, typically yielding 20 s on-source after
a slew/settle time of about 12 s.                                       
We selected the strongest maser channel at an LSR velocity, $V_{\rm LSR}$,
of 56.98 km s$^{-1}$ to serve as the interferometer phase reference in order
to measure the position offsets of the quasars as well as other maser channels.

The data reduction was carried out using the NRAO’s Astronomical
Image Processing System \citep[AIPS]{2003ASSL..285..109G} together with 
a ParselTongue\footnote{\url{https://www.jive.eu/jivewiki/doku.php?id=parseltongue:parseltongue}}
\citep{2006ASPC..351..497K} script. The well-calibrated maser spot
and the background quasars were imaged with the AIPS task IMAGR.        
The background quasars were detected in synthesis maps at all epochs
with SNR $\ge$ 10,                                                      
except for J1853-0010 on 2014 July 22 which was
not detected.                                                       
The position offsets of the quasars relative to the reference maser
spot were derived by fitting Gaussian brightness distributions using
the AIPS task JMFIT.                                                    
We used the AIPS task SAD to fit the positions of all the maser spots.

\section{Results} \label{sec:result} 
\subsection{Spatial Distribution and Internal Motion of Masers} 
\begin{figure}[ht] 
  \centering 
  \includegraphics[width=18cm]{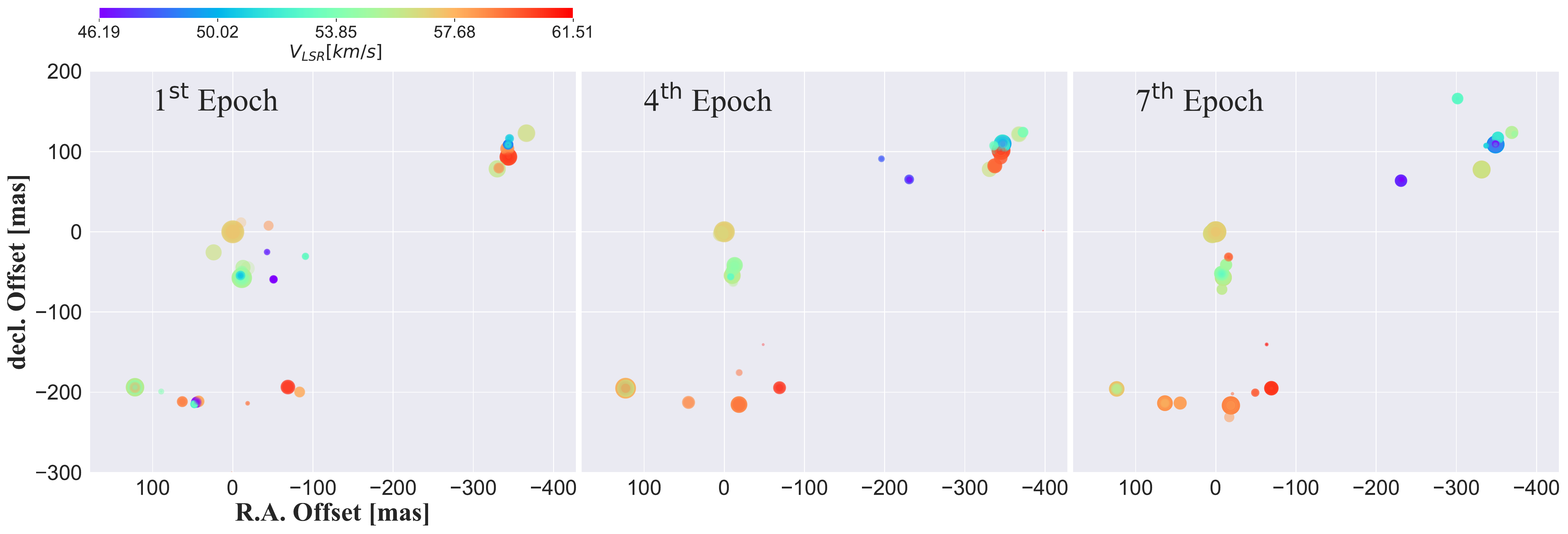} 
  \caption{ 
 Spatial distributions of water-maser emission towards G034.43+0.24
observed at the 1$^{\rm st}$, 4$^{\rm th}$ and 7$^{\rm th}$ epoch.
The V$_{\rm LSR}$ of maser spots are color-coded as indicated
by the color bar, and the logarithm of the brightness is indicated
by the circle size.                                                                   
} 
  \label{fig:maserdis} 
\end{figure}

Figure~\ref{fig:maserdis} shows the spatial distribution of masers
at the 1$^{\rm st}$, 4$^{\rm th}$ and 7$^{\rm th}$ epoch. The masers
are distributed over a region of about 600 by 200 mas elongated in
the SE-NW direction. The general morphology of the distribution remains
stable over 1.5 yrs.                                                    
 
\begin{figure}[H] 
  \centering 
  \includegraphics[scale=0.47]{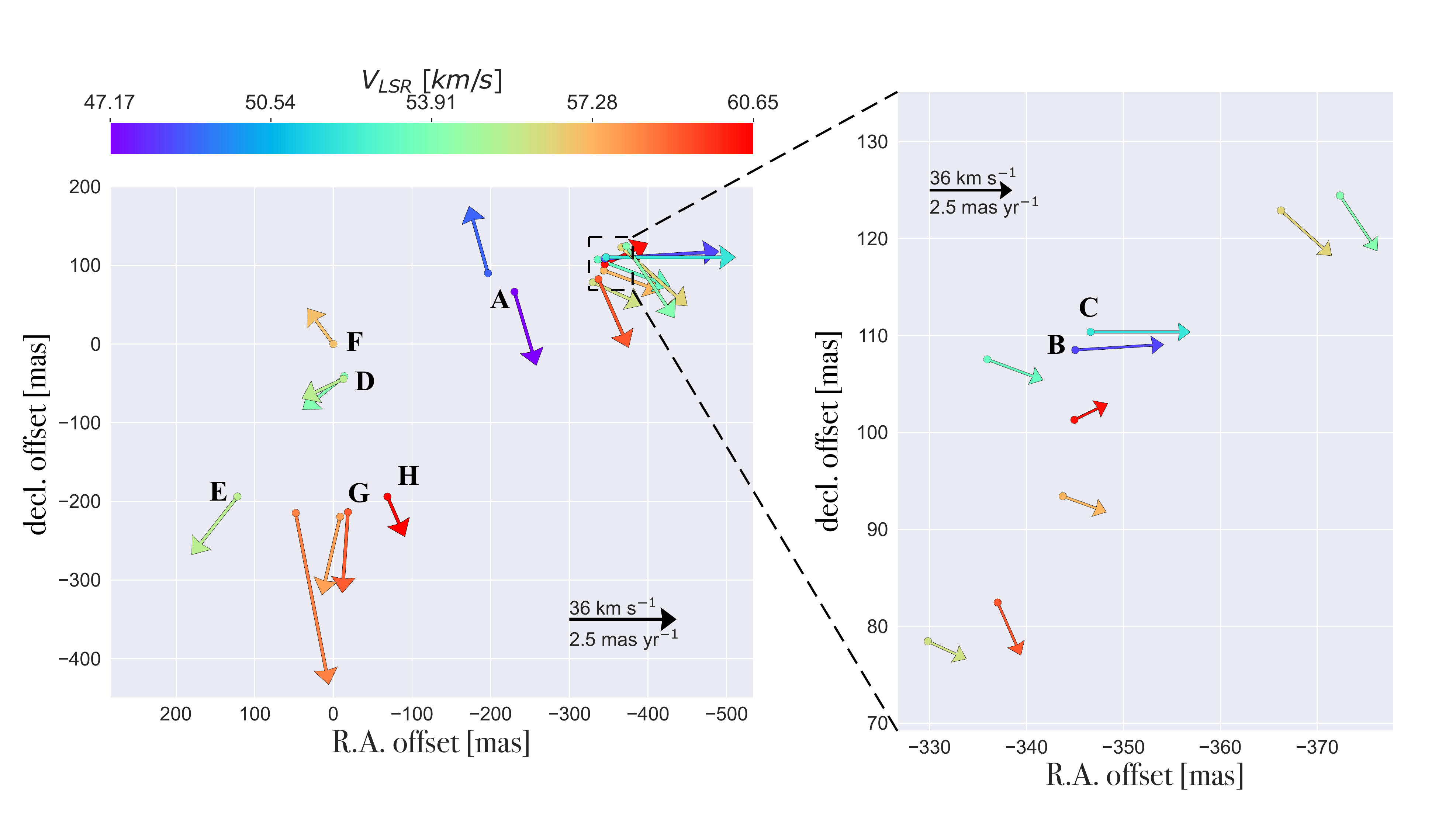} 
  \caption{ 
\emph{Left panel}: The internal proper motions of the maser features.
The labels A through H denote the maser features
used in the parallax determination.  The V$_{\rm LSR}$ of the maser
features are color-coded as indicated by the color bar. The proper
motion amplitude scale is given by the black arrow in the bottom
right corner of the panel. \emph{Right panel}: Zoom-in of the NW
cluster.                                                                
} 
  \label{fig:intmotion} 
\end{figure}

    The internal motions of the water masers can be used to probe
the gas kinematics within G034.43+0.24 MM1. The term ``spot'' used
below refers to an individual maser emission peak in a single velocity
channel, and the term ``feature" refers to a collection of spots
emitting at a similar position. We identified persistent maser spots 
by comparing their positions measured in the different epochs. For
a given spectral channel, if their positions coincided within $\Delta
t~\times$ 5 mas yr$^{-1}$, where $\Delta t$ is the time in years
between epochs, we grouped them for parallax and proper motion fitting. 
For measurement of the internal motions, we grouped
maser spots into a feature if they were detected over three or more
continuous channels.  We used a variance-weighted least-square fit to solve
for a position offset at a reference epoch and a linear motion (relative
to the reference maser spot at \VLSR = 56.98 km s$^{-1}$).
We then calculated the average motion
of all features and subtracted that from each feature's motion. The
internal proper motions are shown in Figure~\ref{fig:intmotion},
ranging from 1.0 to 4.1 mas yr$^{-1}$.                                  
Many of the maser features are moving in the SW-NE direction,
roughly perpendicular to the SE-NW distribution.                        
However, it is worth noting that the average motion subtracted
from all spots is not guaranteed to place the motions in a reference frame of
the central star(s), and one is free to add a single constant vector to all internal
motions.

\subsection{Parallax and Proper Motion Fitting}  \label{sec:fitting}

As discussed in \citet{2012ApJ...744...23Z}, 
a maser ``spot" can be a blend of multiple components, and this
can lead to poor parallax results if the components change strength.
Therefore, we selected spots that appeared not to be blended. 
We identified 47 maser spots belonging to nine 
maser features (labeled A through H, with E composed of two features at
nearly the same position),
which persisted over more than 5 epochs.  
We first performed individual
fits for each spot with respect to each background quasar. 
In order
to account for systematic errors owing to residual atmospheric delays
and potential structural changes in a maser spot and/or a background
quasar, we added ``error floors'' 
in quadrature with the formal Gaussian-fit
uncertainties (which are dominated by thermal noise),
separately for the R.A. and decl. measurements.  The error-floors
were iteratively adjusted until the $\chi^2$ per degree-of-freedom
of the post-fit residuals was close to
unity for both R.A. and decl.

\begin{figure}[ht] 
  \centering 
  \includegraphics[scale=0.4]{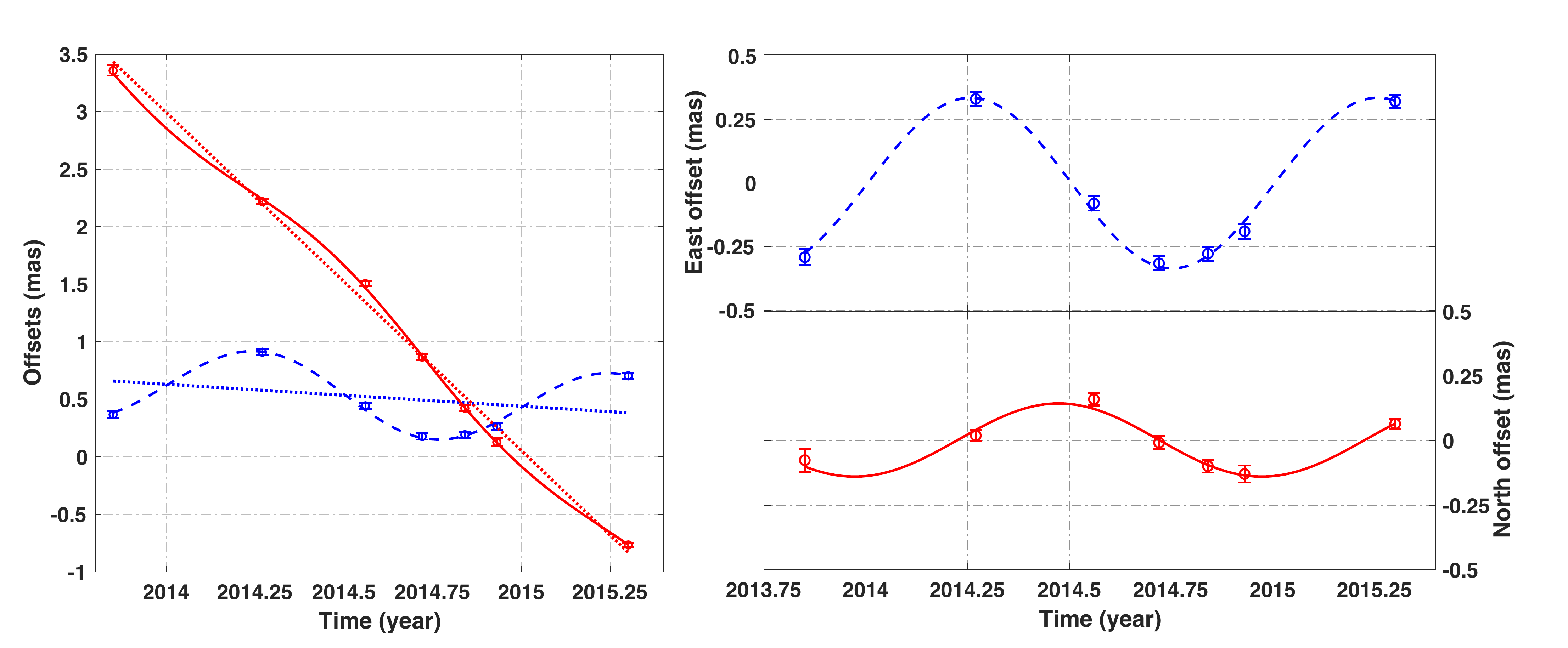} 
  \caption{ 
Parallax and proper motion data (circles with error bars) and best-fitting
models (lines) for the reference maser spot at V$_{\rm LSR}=56.98$ km s$^{-1}$
relative to J1855+0251. \emph{Left panel}:  The red circles are for northward
offsets, and the blue circles are for eastward offsets. The red solid
line and blue dashed line are for the best-fitting models (parallax
+ proper motion). The dotted lines represent the best-fitting proper motions.
\emph{Right panel}: Eastward
(blue) and northward (red) offsets displaying only the parallax signature,
i.e. after removing the best-fitting proper motion.                     
} 
  \label{fig:ppm} 
\end{figure}

\begin{figure}[ht] 
  \centering 
  \includegraphics[width=18cm]{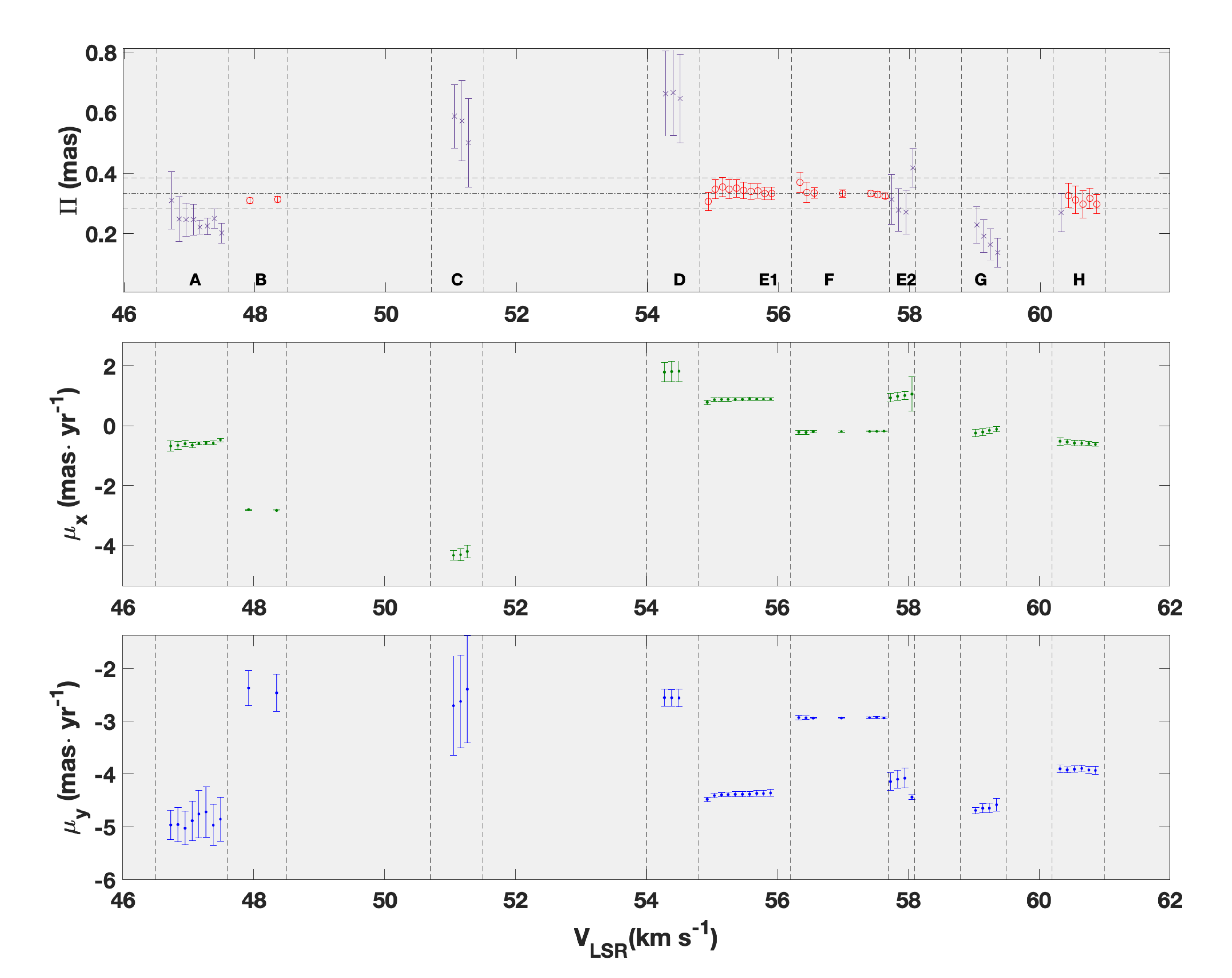} 
  \caption{ 
Individual-channel solutions for parallax and proper motion of the
maser spots in features A through H relative to J1855+0251.
{\it Top Panel:} Red circles indicate maser spots used for our final
estimate of parallax, while purple stars represent the maser spots
not used for parallax.  The dash-dotted line indicates the mean parallax and the
horizontal dashed lines indicate $\pm3\sigma$ for the spots used.
{\it Middle and Bottom Panels:} The fitted eastward and northward proper
motion components.    
} 
  \label{fig:mppm} 
\end{figure}

Figure~\ref{fig:ppm} shows the results
of parallax and proper motion fitting for the reference maser at
\VLSR = 56.98 km s$^{-1}$.  Figure~\ref{fig:mppm} illustrates the
individual solutions of parallax and proper motion for all spots
relative to J1855+0251. While scatter in the proper motions
is expected (owing to internal motions), the parallax values should
be consistent.  As shown in Figure~\ref{fig:mppm}, the parallaxes
of features B, E1\footnote{Maser
feature E1 and E2 appeared at slightly different positions, but are
indistinguishable on a global maser spot map such as Figs.~\ref{fig:maserdis}
and~\ref{fig:intmotion}.}, F, and H are consistent within their uncertainties,
whereas the parallax estimates for features A, C, D,
E2 and G have large uncertainties and do not well agree.  These maser spots
likely suffer from spectral blending of variable
emission components and were not used to estimate the parallax of
G034.43+00.24.

The parallaxes of the remaining 24 maser spots (red open circles
in Figure~\ref{fig:mppm}) agree well.                                   
Therefore, we obtained a single solution, combining these spots, for  
each background quasar, by fitting a single parallax,
while permitting different proper motions for each maser spot.
The background quasar J1853$-$0010 exhibited extended structure
toward the ENE direction, most likely owing to jetted emission, which significantly
degraded the parallax fitting in the R.A. direction.  For this background quasar,
we obtained a parallax estimate of $0.235\pm0.047$ mas, which has twice the formal
uncertainty compared to the unresolved quasars J1855+0251 and J1848+0108.
Since the extended structure seen toward J1853$-$0010 is likely variable, and
thus could systematically affect the parallax estimate, we discarded those data.

For our final result, we only used background quasars J1855+0251 and J1848+0108.
Table~\ref{tab:ppm} lists the combined results for the 24 good maser spots
relative to these background quasars.
We achieved a $\chi^2$ per degree-of-freedom ($\Tilde{\chi}^2$ in Table~\ref{tab:ppm}) near unity
for error floor values of $\pm0.055$ and $\pm0.118$ mas in the R.A. and decl. directions,
respectively.  These indicate realistic single-epoch astrometric accuracies.
We inflated
the parallax uncertainty from the combined fits by a factor of $\sqrt{24}$
in order to account for likely correlations among the positions
of the maser spots caused by uncompensated atmospheric delays between
G034.43+0.24 and a given background quasar.
Then, we (variance-weighted) averaged the parallax results from the two
good background quasars in order to obtain our final result:
an annual parallax of $0.330 \pm 0.018$ mas, corresponding to a distance
of $3.03^{+0.17}_{-0.16}$ kpc.                                          
 
In order to estimate the motion of the central star, we
first averaged the proper motion of spots in a feature and then 
took
the midpoint of the extremes of the motions of all features 
as the best estimate for the motion of the star.  We adopted one-quarter of
the spread between the extremes as a conservative estimate of uncertainty.
This yields an absolute proper motion of $(\mu_x, \mu_y)~=~(-0.96\pm
0.93,-3.38 \pm 0.50)$ mas~y$^{-1}$.                                    
 
\begin{table}[H] \footnotesize
\caption{Parallax and proper motion fits for individual channels and the combined solutions.} 
\begin{tabular}{cccccc} 
\hline 
Background quasar & 
  Feature & 
  \begin{tabular}[c]{@{}c@{}}$V_{LSR}$\\ (km s$^{-1}$)\end{tabular} & 
  \begin{tabular}[c]{@{}c@{}}Parallax\\ (mas)\end{tabular} & 
  \begin{tabular}[c]{@{}c@{}}$\mu_x$\\ (mas y$^{-1}$)\end{tabular} & 
  \begin{tabular}[c]{@{}c@{}}$\mu_y$\\ (mas y$^{-1}$)\end{tabular}
\\ \hline                                                               
J1855+0215 & B        & 47.92 & $0.310\pm 0.008$ & $-2.81\pm 0.02$
& $-2.37\pm 0.33$                \\                                     
           &          & 48.35 & $0.314\pm 0.009$ & $-2.83\pm 0.02$
& $-2.46\pm 0.36$                \\                                     
           & E1       & 54.93 & $0.307\pm 0.031$ & $~~0.78\pm 0.07$
& $-4.48\pm 0.04$                \\                                     
           &          & 55.04 & $0.347\pm 0.032$ & $~~0.87\pm 0.07$
& $-4.41\pm 0.04$                \\                                     
           &          & 55.15 & $0.353\pm 0.032$ & $~~0.88\pm 0.07$
& $-4.39\pm 0.05$                \\                                     
           &          & 55.25 & $0.347\pm 0.031$ & $~~0.88\pm 0.06$
& $-4.39\pm 0.05$                \\                                     
           &          & 55.36 & $0.350\pm 0.029$ & $~~0.88\pm 0.06$
& $-4.38\pm 0.05$                \\                                     
           &          & 55.47 & $0.343\pm 0.027$ & $~~0.89\pm 0.05$
& $-4.38\pm 0.05$                \\                                     
           &          & 55.58 & $0.340\pm 0.026$ & $~~0.89\pm 0.05$
& $-4.38\pm 0.05$                \\                                     
           &          & 55.69 & $0.341\pm 0.023$ & $~~0.89\pm 0.04$
& $-4.37\pm 0.06$                \\                                     
           &          & 55.79 & $0.333\pm 0.022$ & $~~0.89\pm 0.04$
& $-4.37\pm 0.06$                \\                                     
           &          & 55.90 & $0.334\pm 0.021$ & $~~0.89\pm 0.04$
& $-4.36\pm 0.06$                \\                                     
           & F        & 56.33 & $0.370\pm 0.034$ & $-0.22\pm 0.07$
& $-2.93\pm 0.04$                \\                                     
           &          & 56.44 & $0.337\pm 0.033$ & $-0.23\pm 0.07$
& $-2.94\pm 0.04$                \\                                     
           &          & 56.55 & $0.335\pm 0.018$ & $-0.20\pm 0.04$
& $-2.94\pm 0.02$                \\                                     
           &          & 56.98 & $0.337\pm 0.012$ & $-0.19\pm 0.03$
& $-2.94\pm 0.02$                \\                                     
           &          & 57.41 & $0.333\pm 0.011$ & $-0.19\pm 0.02$
& $-2.93\pm 0.02$                \\                                     
           &          & 57.52 & $0.329\pm 0.010$ & $-0.19\pm 0.02$
& $-2.93\pm 0.02$                \\                                     
           &          & 57.63 & $0.324\pm 0.009$ & $-0.18\pm 0.02$
& $-2.94\pm 0.02$                \\                                     
           & H        & 60.43 & $0.327\pm 0.041$ & $-0.54\pm 0.08$
& $-3.92\pm 0.06$                \\                                     
           &          & 60.54 & $0.312\pm 0.045$ & $-0.58\pm 0.09$
& $-3.91\pm 0.06$                \\                                     
           &          & 60.65 & $0.297\pm 0.045$ & $-0.58\pm 0.09$
& $-3.89\pm 0.06$                \\                                     
           &          & 60.76 & $0.317\pm 0.033$ & $-0.59\pm 0.06$
& $-3.92\pm 0.06$                \\                                     
           &          & 60.86 & $0.297\pm 0.031$ & $-0.62\pm 0.06$
& $-3.93\pm 0.07$                \\ \hline                              
           & Combined & $\Tilde{\chi}^2 = 0.91$
& $0.334\pm 0.024$ & $-0.97\pm 0.93$  & $-3.40\pm 0.50$         
\\ \hline                                                               
J1848+0138 & 
  B & 
  47.92 & 
  $0.295\pm 0.009$ & 
  $-2.78\pm 0.02$ & 
  $-2.32\pm 0.32$ \\ 
           &          & 48.35 & $0.307\pm 0.012$ & $-2.82\pm 0.02$
& $-2.38\pm 0.38$ \\                                                    
           & E1       & 54.93 & $0.295\pm 0.030$ & $~~0.81\pm 0.06$
& $-4.45\pm 0.05$ \\                                                    
           &          & 55.04 & $0.338\pm 0.031$ & $~~0.89\pm 0.06$
& $-4.37\pm 0.06$ \\                                                    
           &          & 55.15 & $0.343\pm 0.030$ & $~~0.90\pm 0.06$
& $-4.35\pm 0.06$ \\                                                    
           &          & 55.25 & $0.340\pm 0.029$ & $~~0.90\pm 0.06$
& $-4.35\pm 0.07$ \\                                                    
           &          & 55.36 & $0.345\pm 0.028$ & $~~0.90\pm 0.05$
& $-4.34\pm 0.07$ \\                                                    
           &          & 55.47 & $0.338\pm 0.026$ & $~~0.90\pm 0.05$
& $-4.34\pm 0.07$ \\                                                    
           &          & 55.58 & $0.334\pm 0.026$ & $~~0.92\pm 0.05$
& $-4.34\pm 0.07$ \\                                                    
           &          & 55.69 & $0.334\pm 0.025$ & $~~0.91\pm 0.05$
& $-4.32\pm 0.07$ \\                                                    
           &          & 55.79 & $0.325\pm 0.025$ & $~~0.91\pm 0.05$
& $-4.32\pm 0.07$ \\                                                    
           &          & 55.90 & $0.325\pm 0.025$ & $~~0.92\pm 0.05$
& $-4.31\pm 0.08$ \\                                                    
           & F        & 56.33 & $0.361\pm 0.039$ & $-0.20\pm 0.08$
&  $-2.84\pm 0.07$ \\                                                   
           &          & 56.44 & $0.341\pm 0.028$ & $-0.21\pm 0.06$
&  $-2.87\pm 0.04$ \\                                                   
           &          & 56.55 & $0.328\pm 0.010$ & $-0.17\pm 0.02$
&  $-2.88\pm 0.04$ \\                                                   
           &          & 56.98 & $0.321\pm 0.005$ & $-0.16\pm 0.01$
&  $-2.88\pm 0.04$ \\                                                   
           &          & 57.41 & $0.321\pm 0.005$ & $-0.16\pm 0.01$
&  $-2.87\pm 0.04$ \\                                                   
           &          & 57.52 & $0.317\pm 0.004$ & $-0.16\pm 0.01$
&  $-2.87\pm 0.03$ \\                                                   
           &          & 57.63 & $0.313\pm 0.004$ & $-0.15\pm 0.01$
&  $-2.87\pm 0.03$ \\                                                   
           & H        & 60.43 & $0.310\pm 0.036$ & $-0.51\pm 0.08$
&  $-3.86\pm 0.04$ \\                                                   
           &          & 60.54 & $0.301\pm 0.041$ & $-0.55\pm 0.08$
&  $-3.85\pm 0.05$ \\                                                   
           &          & 60.65 & $0.290\pm 0.040$ & $-0.56\pm 0.08$
&  $-3.83\pm 0.05$ \\                                                   
           &          & 60.76 & $0.308\pm 0.028$ & $-0.56\pm 0.05$
&  $-3.86\pm 0.05$ \\                                                   
           &          & 60.86 & $0.289\pm 0.025$ & $-0.60\pm 0.05$
&  $-3.87\pm 0.06$ \\ \hline                                            
           & Combined & $\Tilde{\chi}^2 = 0.97$
& $0.326\pm0.027$  & $-0.95\pm 0.93$  &  $-3.35\pm 0.50$        
\\ \hline                                                               
Average    &          &       & $0.330\pm0.018$  & $-0.96\pm 0.93$
&  $-3.38\pm 0.50$\\                                                    
\hline 
\end{tabular} 
\tablecomments{ The ``Combined" fits solved for a single parallax parameter; see the text for an explanation of the ``Combined" proper motion estimates. The ``Average" parallax is a variance-weighted average of the two background quasars results.  
}
\label{tab:ppm} 
\end{table}

\subsection{3-dimensional kinematic distance} 
\label{sec:3d} 
In order to provide an independent check on our parallax distance,
we used our measured proper motion and \VLSR~to derive
a 3-dimensional (3D) kinematic distance estimate.                            
The 3D kinematic distance method has been
described and evaluated by \citet{2022AJ....164..133R}. 
From Bayes' theorem,
the posterior distribution function (PDF) for distance $d$ can be constructed
from likelihood functions from measurements of \VLSR~and proper motions in
Galactic longitude, $\mu_l$, and latitude, $\mu_b$
(calculated from R.A. and decl. motions): 
$P(d|V_{\rm LSR},\mu_l,\mu_b,RC)\propto P(V_{\rm LSR},\mu_l,\mu_b|d,RC)\times P(d)$,
where RC indicates an assumed Galactic rotation curve and
the prior distance probability function can be taken to be flat.
The likelihood functions for the $V_{\rm LSR}$ and either proper motion component
are given by the following:                                                              
\begin{equation} 
    P(d|V_{\rm LSR}, \sigma_{V})\propto \frac{1}{\sigma_V}e^{-\Delta
V^2/2\sigma_{V}^2},                                           
\end{equation} 
and
\begin{equation} 
    P(d|\mu,\sigma_{\mu})\propto \frac{1}{\sigma_{\mu}}e^{-\Delta\mu^2/2\sigma_{\mu}^2},
\end{equation} 
where $\Delta V$ and $\Delta\mu$ are the differences between the measured and
the model values using equations given by \citet{2011PASJ...63..813S}, and 
$\sigma_V$ and $\sigma_\mu$ are uncertainties in the measurements.                       
Multiplying the component likelihoods leads to the 3D kinematic distance PDF.

\begin{figure}[H] 
  \centering 
  \includegraphics[scale=0.6]{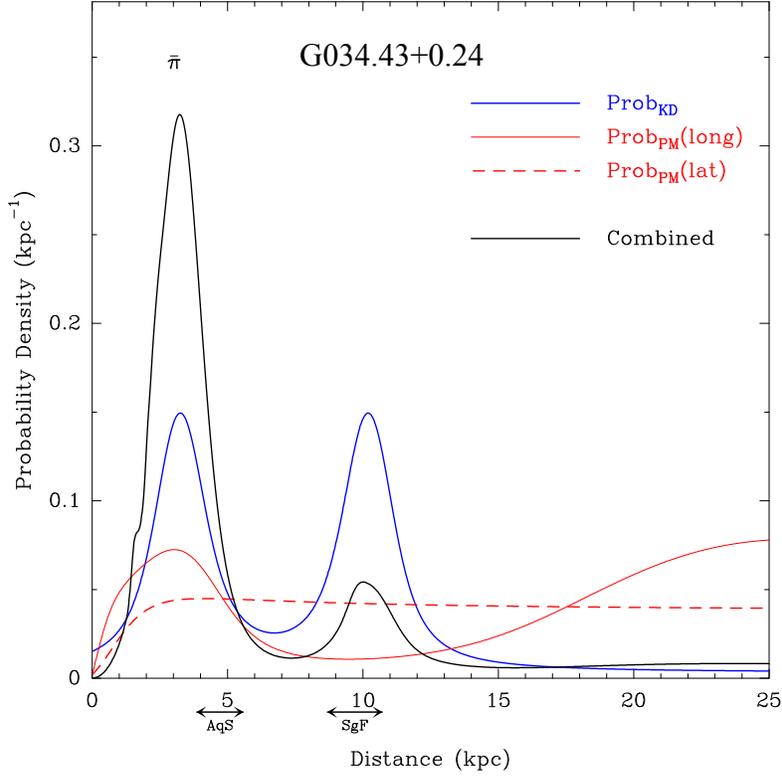} 
  \caption{ 
    Likelihood functions for G034.43+0.24 as a function of distance
for radial velocity (blue line), proper motion in Galactic longitude
(red line), and Galactic latitude (red dashed line). The black
posterior PDF combines the three component likelihoods and yields
a 3D kinematic distance estimate of $3.2\pm0.7$ kpc.  The $\pi$ symbol
indicates our parallax distance, and the line above it gives its
$\pm1\sigma$ uncertainty.      
} 
  \label{fig:distance} 
\end{figure}

We adopted an LSR velocity of $55\pm5$ km s$^{-1}$ for G034.43+00.24
which falls within the range of the water masers features and
is close to that reported by \citet{2006ApJ...653.1325S} for CO emission.
Our proper motion measurements in equatorial coordinates transform to
$\mu_l=-3.45\pm0.86$ and $\mu_b=-0.69\pm0.57$ mas y$^{-1}$.
When evaluating the likelihoods we used the Galactic rotation curve
of~\citet{2019ApJ...885..131R}, based on parallax distances and proper
motions of $\approx150$ massive young stars.
Fig.~\ref{fig:distance} shows likelihoods
as a function of distance for the LSR velocity, yielding distance
estimates of 3.3 or 10.2 kpc (with the standard 1-dimensional kinematic
distance ambiguity), as well as for the proper motion in Galactic
longitude and latitude.  Note that the likelihood for the longitude motion
strongly favors the near distance.
Combining the individual likelihoods yields a 3D kinematic distance
of $3.2\pm0.7$ kpc for G034.43+0.24.  This is fully consistent with our
parallax measurement.

With our 6-dimensional phase-space measurements, we can calculate
the full 3-dimensional velocity vector of G034.43+00.24.   Then,
subtracting circular motion at its location in the Galaxy using the
rotation curve of \citet{2019ApJ...885..131R}, we found the peculiar
(non-circular) motion components
\footnote{The uncertainties in the peculiar motions of G034.43+0.24 include the uncertainties in its distance, proper motions, and the underlying Galactic model of \citet{2019ApJ...885..131R}.}  
to be                                  
$U_p=1\pm10$, $V_p=2\pm11$, and $W_p=-2\pm8$ \kms, 
where $U_p$ is toward the Galactic center at the location of the source, 
$V_p$ is in the direction of Galactic rotation, and $W_p$ is toward
the North Galactic Pole.                                   
Thus, the star(s) exciting the water masers in G034.43+00.24 are in a 
nearly circular Galactic orbit, characteristic of massive young stars. 
 
\section{Discussion} \label{sec:discuss} 
 
\subsection{Comparison with previous results} 
 
For most sources with parallax distances measured by both the VERA
project \citep[see][]{2020PASJ...72...50V} and the BeSSeL Survey
\citep[see][]{2019ApJ...885..131R}, there is good agreement, i.e.,
the differences are generally within $\pm2$ times their joint uncertainty.
However, our measured distance for G034.43+0.24 of 3.03$^{+0.17}_{-0.16}$
kpc differs from the VERA result of $1.56^{+0.12}_{-0.11}$ kpc measured
by~\citet{2011PASJ...63..513K} by nearly eight times their joint
uncertainty.                                                            
While we are confident in our result, we can
only offer some speculations as to why the~\citet{2011PASJ...63..513K}
parallax distance could be significantly different.  Firstly, G034.43+0.24
has a Declination very close to zero, which can be problematic for
the VERA array with only 4 antennas~\citep{2011PASJ...63..513K}.
Secondly, the Kurayama observations relied on a single background
quasar. If that quasar experienced changes in structure over the
time span of their observations, this could affect the parallax measurement
(as we found for quasar J1853$-$0010),
especially when only the R.A. component was measured in their case.   
 
\subsection{Water masers tracing dense gas bubbles} 
\begin{figure}[H] 
  \centering 
  \includegraphics[width=18cm]{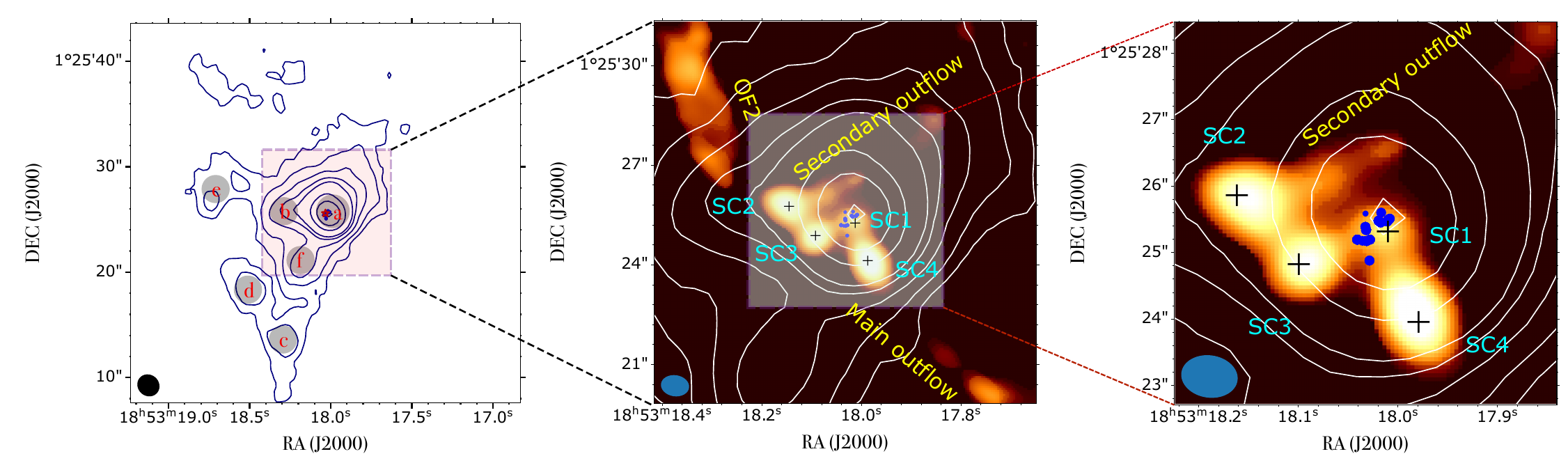} 
  \caption{ 
Maps of the IRDC G034.43+0.24. \emph{Left panel}: ATOMS continuum
emission at 3 mm. The labels with grey circles represent the cores
identified in \citet{2022MNRAS.510.5009L}. The red star denotes the
position of the maser clump. The restoring beam (black) is indicated
in the bottom left corner. \emph{Middle panel}: zoom-in contour of
the region of MM1-a overlaid on CO $J=3-2$ emission distribution
integrated between +70 and +75 km s$^{-1}$ from \citet{2021A&A...649A.139I}
to emphasize the fragmentation. The blue circles indicate individual
maser features. The restoring beam is indicated in the bottom left
corner. \emph{Right panel}: zoom-in image of the \emph{Middle panel}.}  
  \label{fig:origin} 
\end{figure}

    The absolute position of the reference maser spot at a \VLSR\
of 56.98 km s$^{-1}$ is $\alpha_{\rm J2000}=18^{\rm h}53^{\rm m}18
\fs 0326$, $\delta_{\rm J2000} = +01^{\circ}25^{\prime}25\farcs 4121$.
\footnote{The absolute position was determined by referencing the background quasar J1855+0251 located at $\alpha_{\rm J2000}=18^{\rm h}55^{\rm m}35
\fs 4364$, $\delta_{\rm J2000} = +02^{\circ}51^{\prime}19\farcs 5630$.}
By comparing the absolute position with the ALMA image by \citet{2022MNRAS.510.5009L}
and \citet{2021A&A...649A.139I}, as shown in Figure~\ref{fig:origin},
the water maser clumps are associated in projection with subcore
1 (SC1) of MM1-a of IRDC G034.43+0.24.                                  
 
    Water masers are seen in outflows and may be excited by shocks,
e.g., J type~\citep{1989ApJ...346..983E} or C  type~\citep{1996ApJ...456..250K},
when outflowing gas interacts with ambient material. It is noteworthy
that the internal proper motions of the maser features range from
14 to 58 km s$^{\rm -1}$ (derived from 1.0 to 4.1 mas yr$^{-1}$
at the distance of 3.03 kpc), while the range of V$_{\rm LSR}$ values
is only about 13 km s$^{\rm -1}$. This asymmetry could occur if the masers
are in a thin expanding shell, where tangential amplification lengths
are the longest. To interpret the observed maser spatial-kinematics,
we propose that the masers are tracing an expanding gas shell. However,
as  revealed by the ALMA observations of 
\citet{2022MNRAS.510.5009L}, on larger scales than our masers the
outflow activity is complex, including
four outflow lobes launched from MM1-a, and suggesting that MM1-a may
host multiple protostars.                                               
    As shown in Figures~\ref{fig:maserdis} and \ref{fig:intmotion},
masers seem to assemble in a long arc, or more likely two arc-like
clusters in the SE (R.A. offset $>-100$ mas) and NW (R.A. offset
$<-100$ mas), moving toward the south and southwest, respectively.        
In addition, in Figure~\ref{fig:intmotion}, we notice a radial
velocity difference between the two clusters, which may suggest that
they are excited by different protostars.
 
    Considering the high degree of stellar multiplicity in massive
star-formation regions, it is also possible that the water masers
are tracing two dense gas bubbles compressed by outflows launched
from multiple protostars in MM1-a.  This is similar to the water masers in 
G011.92$−$0.61 and G035.02$+$0.35 reported by \citet{2019A&A...631A..74M},
which show inhomogeneous spatial distributions, separating into two
clusters at different velocities, also suggesting that two YSOs excite
different water maser clusters.                                         
Observations with higher resolution of at least 0.3$^{\prime\prime}$
(half of the spread of the maser clusters) toward MM1-a using ALMA might
be able to resolve multiple protostars.

\subsection{Spiral arm association} 
 
\begin{figure}[H] 
  \centering 
  \includegraphics[scale=0.5]{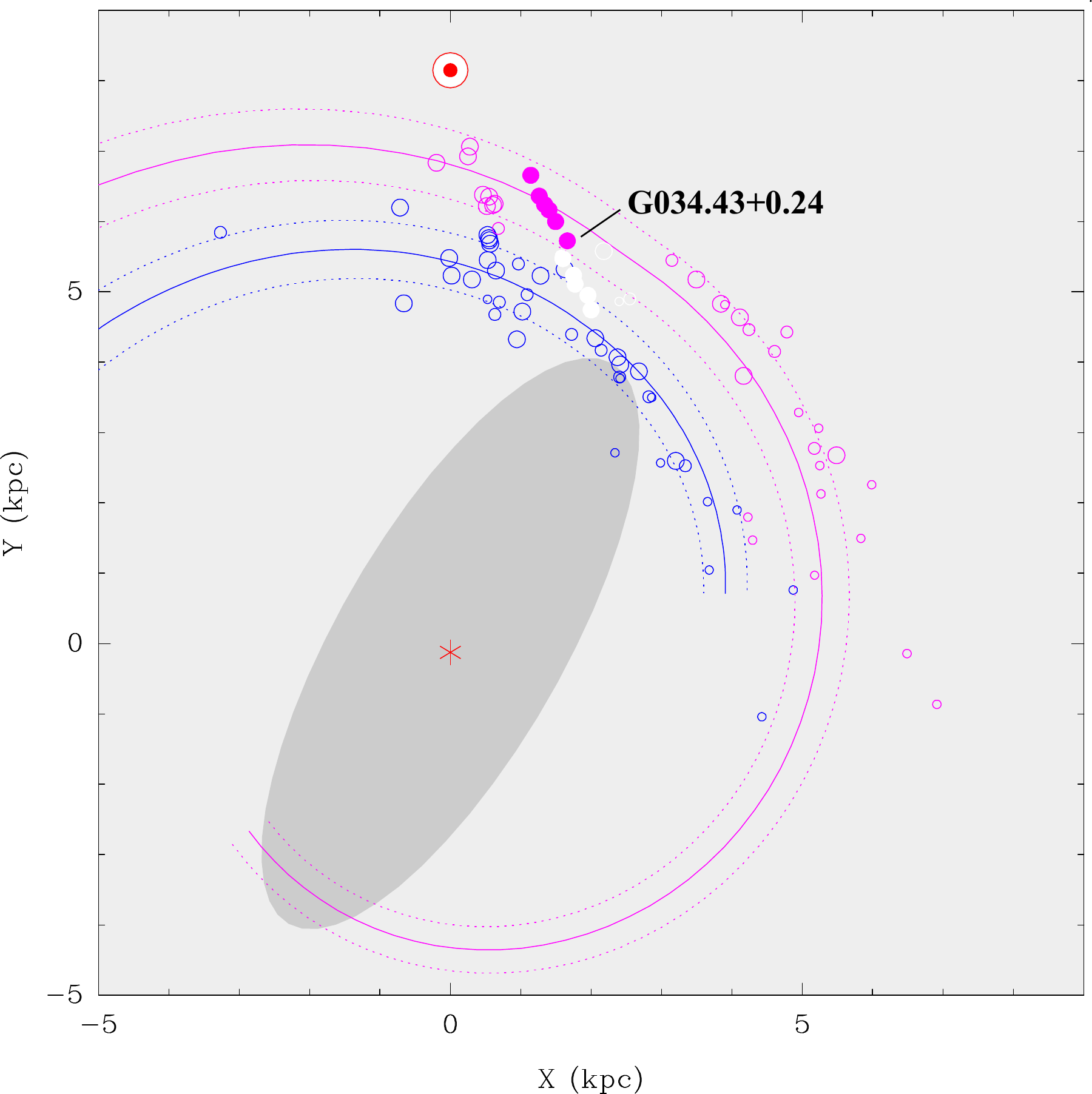} 
  \caption{ 
Plan-view map of the Milky Way as viewed from the North Galactic Pole and adapted
from \citet{2019ApJ...885..131R}. Here we only show the Sagittarius–Carina
arm (purple lines and symbols) and Scutum–Centaurus arm (blue lines
and symbols).  Circles locate massive young stars with maser parallaxes.
The filled purple circles identify the linear arrangement
of five masers listed in Table~\ref{tab:linear} and one 
possibly unrelated source (G037.43+1.52). The white filled
circles are sources associated with a spur between the Sagittarius and Scutum
arms.  The Sun is indicated by the sun symbol at (0, 8.15) kpc and the Galactic Center
is at (0, 0) kpc.  The shaded ellipse schematically represents the long Galactic Bar.
} 
  \label{fig:mwplane} 
\end{figure}

\begin{figure}[H] 
  \centering 
  \includegraphics[scale=0.6]{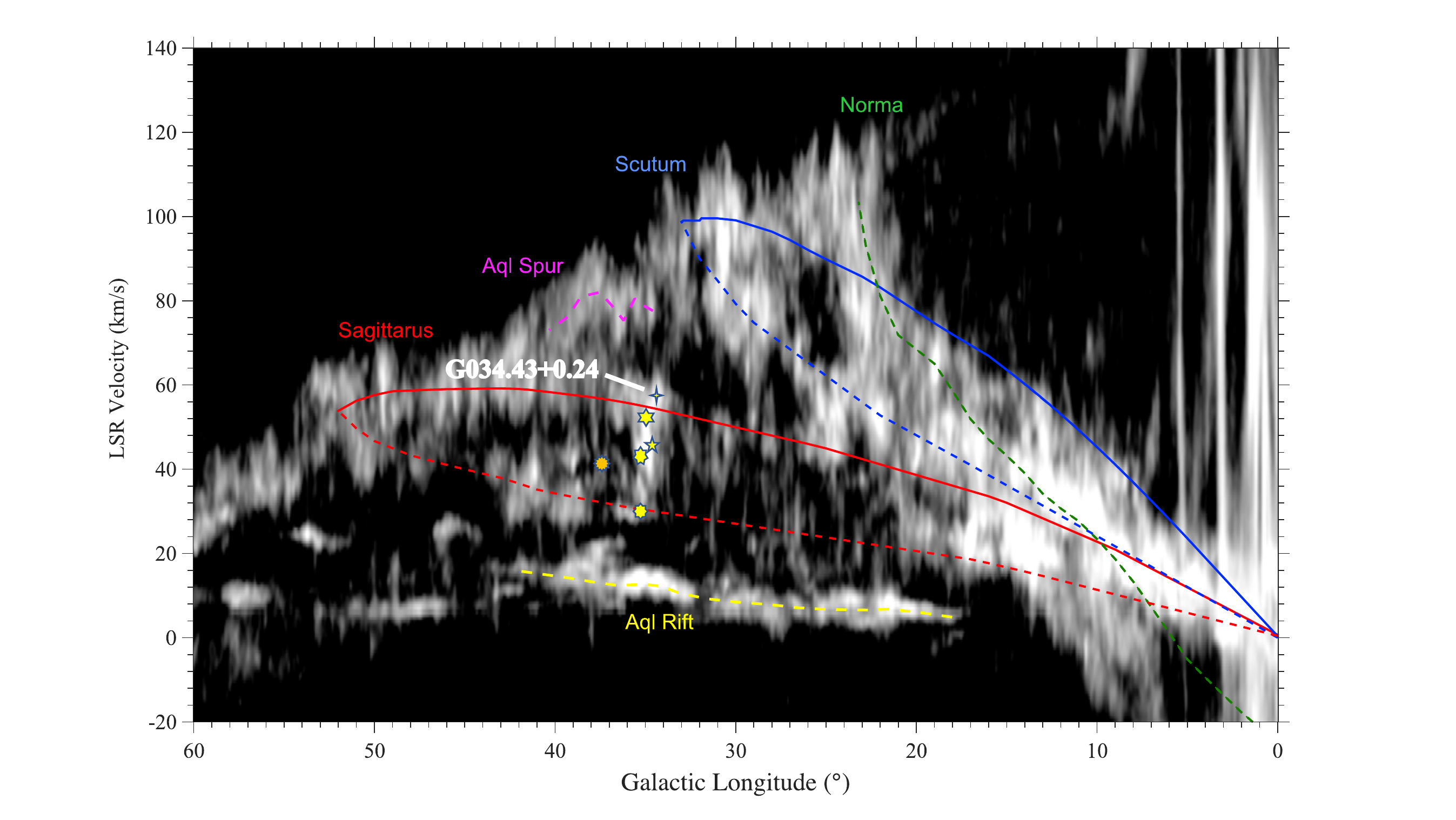} 
  \caption{ 
Longitude--Velocity traces of spiral arms superposed on CO emission
after Fig. 7 of \citet{2016ApJ...823...77R}.  The near and far sides of the arm
are indicated with dashed and solid lines.
The four-pointed star is G034.43+0.24, the pentagram is G034.79--1.38, the
hexagram is G035.02+0.34, the heptagram is G035.20--1.73, and the octagram
is G035.20--0.74.  The orange dodecagram is G037.43+1.52.
See Table~\ref{tab:linear} for source distances and kinematics.
} 
  \label{fig:lv} 
\end{figure}

    Our parallax, confirmed by its 3D kinematic distance, places G034.43+0.24
in the Sagittarius spiral arm as traced by massive young stars with
maser parallaxes \citep{2019ApJ...885..131R}.
At first glance this 
is surprising, as its \VLSR\ of 55 \kms\ would favor placing it in
the far portion of the Sagittarius spiral arm, well past the arm tangency,
based on the longitude--velocity ($l-V$) trace of the spiral arm model shown in
Fig.~\ref{fig:lv}.  However, G034.43+0.24 appears at one end of a ``vertical''
structure in the CO $l-V$ emission, which bridges the near and
far sides of the Sagittarius arm in that plot.  
Sources in that $l-V$ bridge also display a linear pattern on a ($X,Y$) map of
the Milky Way (see filled purple circles in Fig.~\ref{fig:mwplane}).
\footnote{A spurious linear structure in a map which points back
to the Sun can be created from a group of sources at similar distances
but with significant distance errors (so-called ``fingers of God'').
However, as can be seen in Fig.~\ref{fig:mwplane},
the linear arrangement is slightly misaligned and does not directly point back
to the Sun. Also, the parallax-distance errors are not large enough to
cause the observed linear arrangement. Finally, since the sources also trace
an unusual structure in $l-V$ space, we reject the ``fingers of God'' interpretation.}

Table~\ref{tab:linear} lists parallax distances and full 3D kinematic information
for the sources in the bridge.  We note that the non-circular motions of
all of these sources are small ($<10$ \kms), which argues against a
large-scale structure formed by the stellar wind of the precursor O-type star
associated with supernova remnant W44 \citep[see][]{1985ApJ...288L..17R}.
These sources span the extent of the Sagittarius arm at a pitch
angle of $\approx45^\circ$; were the arrangement to extend outside
the arm it could be considered a spur, such as the white filled circles in
Fig.~\ref{fig:mwplane} between the Scutum and Sagittarius spiral arms.
However, being inside the arm they may trace an internal filamentary structure,
possibly similar to the ``Radcliffe wave" seen inside the Local spiral arm
\citep{2020Natur.578..237A}.

\begin{table}[H] 
\caption{Massive young stars in the linear arrangement.} 
\setlength\tabcolsep{-5pt} 
\begin{tabular}{lccccccc} 
\hline 
\begin{tabular}[c]{@{}c@{}}Source\\ \end{tabular} & 
\begin{tabular}[c]{@{}c@{}}Distance\\ (kpc)\end{tabular} & 
\begin{tabular}[c]{@{}c@{}}V$_{\rm LSR}$\\ (km s$^{-1}$)\end{tabular} & 
\begin{tabular}[c]{@{}c@{}}$\mu_x$\\ (mas y$^{-1}$)\end{tabular}     & 
\begin{tabular}[c]{@{}c@{}}$\mu_y$\\ (mas y$^{-1}$)\end{tabular} & 
\begin{tabular}[c]{@{}c@{}}$U_p$\\ (km s$^{-1}$)\end{tabular} & 
\begin{tabular}[c]{@{}c@{}}$V_p$\\ (km s$^{-1}$)\end{tabular} & 
\begin{tabular}[c]{@{}c@{}}$W_p$\\ (km s$^{-1}$)\end{tabular} \\
\hline 
 G035.20$-$0.74 & ~~~~~~~~~2.19$^{+0.24}_{-0.20}$ & ~~~~~~~~~30\,$\pm$\,7
& ~~~~~~~--0.18\,$\pm$\,0.50 & ~~~~~--3.63\,$\pm$\,0.50 & ~~~~~--4\,$\pm$\,7
& ~~~~~--8\,$\pm$\,7 & ~~~~~--8\,$\pm$\,6 \\                            
 G035.02$+$0.34 & ~~~~~~~~~2.33$^{+0.24}_{-0.20}$ & ~~~~~~~~~52\,$\pm$\,5
& ~~~~~~~--0.92\,$\pm$\,0.90 & ~~~~~--3.61\,$\pm$\,0.90 & ~~~~~12\,$\pm$\,9
& ~~~~~~5\,$\pm$\,8   & ~~~~~~--1\,$\pm$\,10\\                          
 G035.20$-$1.73 & ~~~~~~~~~2.40$^{+0.07}_{-0.06}$ & ~~~~~~~~~43\,$\pm$\,3
& ~~~~~~~--0.68\,$\pm$\,0.44 & ~~~~~--3.60\,$\pm$\,0.44 & ~~~~~~3\,$\pm$\,4
& ~~~~~--2\,$\pm$\,4 & ~~~~~--5\,$\pm$\,5\\                             
 G034.79$-$1.38 & ~~~~~~~~~2.62$^{+0.15}_{-0.13}$ & ~~~~~~~~~45\,$\pm$\,5
& ~~~~~~~--0.31\,$\pm$\,0.62 & ~~~~~--2.80\,$\pm$\,0.72 & ~~~~~--7\,$\pm$\,8
& ~~~~~~4\,$\pm$\,7   & ~~~~~--6\,$\pm$\,8\\                            
 G034.43$+$0.24 & ~~~~~~~~~3.03$^{+0.17}_{-0.16}$ & ~~~~~~~~~55\,$\pm$\,5
& ~~~~~~~--0.96\,$\pm$\,0.93 & ~~~~~--3.38\,$\pm$\,0.50 & ~~~~~~1\,$\pm$\,10
& ~~~~~~2\,$\pm$\,11   & ~~~~~~--2\,$\pm$\,8\\                          
\hline 
\end{tabular} 
\label{tab:linear} 
~\\ 
~\\ 
\textbf{Notes}: Columns from left to right give Galactic coordinates,
parallax distances, local standard-of-rest velocities, eastward and northward
proper motion components, and peculiar (non-circular) motion components toward
the Galactic center, in the direction of Galactic rotation and toward the North
Galactic Pole.  The peculiar motions are relative to the rotation curve of
\citet{2019ApJ...885..131R}
The astrometric parameters of G035.20$-$0.74, G035.02$+$0.34, G035.20$-$1.73,
and G034.79$-$1.38 are taken from \citet{2019ApJ...885..131R}.          
\end{table} 
 
\section{Summary} 
\label{sec:sum} 
We have measured an annual parallax of 0.330~$\pm$~0.018 mas for the
H$_2$O masers in the massive star-forming region G034.43+0.24, using
VLBA observations as a part of the BeSSeL survey.                                         
The distance to the source is 3.03$^{+0.17}_{-0.16}$ kpc. We also
estimate a 3D kinematic distance of 3.2 $\pm$ 0.7 kpc,
which agrees well with the parallax result.
The parallax distance places G034.43+0.24 near
the inner edge of the Sagittarius arm and at one end of a linear
arrangement of massive star-formation regions, possibly tracing
a filamentary structure like the ``Radcliffe wave'' in the Local arm.   
 
Compared to previous ALMA observations, the water masers are associated
in projection with IRDC G034.43+0.24 MM1-a which hosts multiple protostars.
The masers are spread over a region of 600 by 200 mas elongated in the
SE-NW direction. The masers likely assemble into two arc-like structures,
with the southeast arc moving toward the south and the northwest arc moving
southeast. We suggest
that the arc-like structures may represent expanding dense gas
bubbles compressed by outflows launched from different protostars.

~\\ 
\\ 
This  work  is partly supported by the Natural Science Foundation
of China (NSFC, grant No. U1831136, U2031212, and 11903079) and Shanghai
Astronomical Observatory (N-2020-06-09-005).  We thank Tie Liu 
for providing the ATOMS data.

\vspace{5mm} 
\facilities{VLBA, ALMA}

\software{astropy \citep{2022ApJ...935..167A}, 
          aplpy \citep{2019zndo...2567476R}, 
          AIPS \citep{2003ASSL..285..109G}, 
          ParselTongue \citep{2006ASPC..351..497K} 
          }

\newpage 
\bibliography{main_mjr}{} 
\bibliographystyle{aasjournal}

\appendix 
 
\section{SUPPLEMENTAL MATERIAL} 
\begin{table*}[hbt!] 
\setlength\tabcolsep{-3pt} 
\caption{Identified maser features and internal proper motions.} 
\begin{tabular}{lcccccc} 
\hline 
Feature & 
\begin{tabular}[c]{@{}c@{}}V$_{\rm LSR}$\\ (km s$^{-1}$)\end{tabular} & 
\begin{tabular}[c]{@{}c@{}}$\Delta \alpha$\\ (mas)\end{tabular} & 
\begin{tabular}[c]{@{}c@{}}$\Delta \delta$\\ (mas)\end{tabular} & 
\begin{tabular}[c]{@{}c@{}}$\mu_{\rm x}$\\ (mas yr$^{-1}$)\end{tabular}
&                                                                       
\begin{tabular}[c]{@{}c@{}}$\mu_{\rm y}$\\ (mas yr$^{-1}$)\end{tabular}
&                                                                       
\begin{tabular}[c]{@{}c@{}}Brightness\\ (Jy beam$^{-1}$)\end{tabular}
\\                                                                      
\hline 
 1(E)  &  ~~~~~~~55.47 &~~~~~~~  121.98 &~~~~~~~ -193.84 &~~~~~~~
1.07 &~~~~~~~ -1.36 &~~~~~~~  61.3 \\                                   
 2(H)  &  ~~~~~~~60.65 &~~~~~~~  -68.69 &~~~~~~~ -194.04 &~~~~~~~
-0.41 &~~~~~~~ -0.94 &~~~~~~~  12.7 \\                                  
 3(G)  &  ~~~~~~~59.03 &~~~~~~~  -18.50 &~~~~~~~ -213.84 &~~~~~~~
0.13 &~~~~~~~ -1.88 &~~~~~~~   1.3 \\                                   
 4     &  ~~~~~~~57.73 &~~~~~~~   -8.53 &~~~~~~~ -219.55 &~~~~~~~
0.42 &~~~~~~~ -1.83 &~~~~~~~   3.6 \\                                   
 8     &  ~~~~~~~58.36 &~~~~~~~   47.92 &~~~~~~~ -214.80 &~~~~~~~
-0.77 &~~~~~~~ -4.01 &~~~~~~~   2.3 \\                                  
 9(F)  &  ~~~~~~~56.98 &~~~~~~~    0.00 &~~~~~~~    0.00 &~~~~~~~
0.61 &~~~~~~~  0.83 &~~~~~~~ 560.3 \\                                   
 10    &  ~~~~~~~54.06 &~~~~~~~  -13.96 &~~~~~~~  -40.99 &~~~~~~~
0.97 &~~~~~~~ -0.78 &~~~~~~~   0.9 \\                                   
 11(D)    &  ~~~~~~~55.47 &~~~~~~~  -12.69 &~~~~~~~  -44.45 &~~~~~~~
0.97 &~~~~~~~ -0.46 &~~~~~~~  16.2 \\                                   
 12    &  ~~~~~~~48.94 &~~~~~~~ -196.45 &~~~~~~~   89.97 &~~~~~~~
0.43 &~~~~~~~  1.57 &~~~~~~~   0.9 \\                                   
 13(A) &  ~~~~~~~47.17 &~~~~~~~ -230.25 &~~~~~~~   66.25 &~~~~~~~
-0.51 &~~~~~~~ -1.72 &~~~~~~~   0.5 \\                                  
 14    &  ~~~~~~~56.01 &~~~~~~~ -329.81 &~~~~~~~   78.44 &~~~~~~~
-1.17 &~~~~~~~ -0.53 &~~~~~~~  37.1 \\                                  
 15    &  ~~~~~~~59.14 &~~~~~~~ -337.01 &~~~~~~~   82.44 &~~~~~~~
-0.70 &~~~~~~~ -1.60 &~~~~~~~   3.7 \\                                  
 16    &  ~~~~~~~57.19 &~~~~~~~ -343.74 &~~~~~~~   93.42 &~~~~~~~
-1.32 &~~~~~~~ -0.48 &~~~~~~~   3.5 \\                                  
 17    &  ~~~~~~~60.43 &~~~~~~~ -344.94 &~~~~~~~  101.29 &~~~~~~~
-1.01 &~~~~~~~  0.50 &~~~~~~~  15.4 \\                                  
 18    &  ~~~~~~~53.31 &~~~~~~~ -335.95 &~~~~~~~  107.54 &~~~~~~~
-1.70 &~~~~~~~ -0.63 &~~~~~~~   4.6 \\                                  
 19(B) &  ~~~~~~~48.35 &~~~~~~~ -345.03 &~~~~~~~  108.51 &~~~~~~~
-2.67 &~~~~~~~  0.17 &~~~~~~~   0.9 \\                                  
 20(C) &  ~~~~~~~52.02 &~~~~~~~ -346.60 &~~~~~~~  110.37 &~~~~~~~
-3.02 &~~~~~~~ 0.00 & ~~~~~~~ 15.3 \\                                   
 21    &  ~~~~~~~56.44 &~~~~~~~ -366.26 &~~~~~~~  122.90 &~~~~~~~
-1.53 &~~~~~~~ -1.37 &~~~~~~~  45.5 \\                                  
 22    &  ~~~~~~~54.07 &~~~~~~~ -372.35 &~~~~~~~  124.45 &~~~~~~~
-1.13 &~~~~~~~ -1.68 &~~~~~~~   3.3 \\                                  
\hline 
\end{tabular} 
\label{tab:propermotion} 
~\\ 
~\\ 
\textbf{Notes}: The letters A-H in parentheses are the same as Fig.~\ref{fig:intmotion},
Fig.~\ref{fig:mppm} and Table~\ref{tab:ppm}.  V$_{\rm LSR}$ is for
the peak channel of a maser feature.  R.A. and decl. offset represent
the position offsets relative to the position of the reference maser
at $\alpha_{\rm J2000}=18^{\rm h}53^{\rm m}18 \fs 0326$, $\delta_{\rm
J2000} = +01^{\circ}25^{\prime}25\farcs 4121$.  Brightness is the peak
at the epoch that a maser feature first appeared.      
\end{table*} 
\clearpage

\end{document}